\documentclass[sigconf]{acmart} 

\usepackage{tabularx}
\usepackage{booktabs}
\usepackage{array}
\usepackage{graphicx}
\usepackage{subcaption}
\usepackage{multirow}
\usepackage{adjustbox}

\usepackage{framed}
\usepackage[strict]{changepage}
\usepackage{subcaption}
\usepackage{placeins}
\usepackage{lettrine} 
\definecolor{formalshade}{rgb}{0.95,0.95,1}
\definecolor{darkblue}{rgb}{0.36,0.54,0.66}

\newenvironment{formal}{%
  \MakeFramed{\advance\hsize-\width\FrameRestore}%
  \noindent\hspace{-4.55pt}
  \begin{adjustwidth}{}{7pt}%
  \vspace{2pt}\vspace{2pt}%
}
{%
  \vspace{2pt}\end{adjustwidth}\endMakeFramed%
}

\AtBeginDocument{%
  }

\copyrightyear{2026}
\acmYear{2026}
\setcopyright{cc}
\setcctype{by}
\acmConference[HAI '26]{Proceedings of the 14th International Conference on Human-Agent Interaction}{November 16--19, 2026}{Osaka, Japan}
\acmBooktitle{Proceedings of the 14th International Conference on Human-Agent Interaction (HAI '26), November 16--19, 2026, Osaka, Japan}
\acmDOI{10.1145/3841580.3841625}
\acmISBN{979-8-4007-2575-3/2026/11}

\begin{document}

\title[Delegating or Doing?]{Delegating or Doing? Understanding User Behavior\\ in Hybrid Human–Agent Interfaces}

\author{Gavin Dizon}
\orcid{0009-0009-8057-7072}
\affiliation{%
  \institution{Future University, Hakodate}
  \city{Hakodate}
  \state{Hokkaido}
  \country{Japan}
}
\email{g-dizon@sumilab.org}

\author{Tyrone Justin {Sta. Maria}}
\orcid{0000-0002-6826-7890} 
\affiliation{%
  \institution{De La Salle University}
 \city{Manila}
  \country{Philippines}}
  \email{tyrone_stamaria@dlsu.edu.ph}

\author{Jordan Aiko {Deja}}
\authornote{Equal Senior contribution}
\orcid{0000-0001-9341-6088}
\affiliation{%
  \institution{De La Salle University}
  \city{Manila}
  \country{Philippines}}
  \email{jordan.deja@dlsu.edu.ph}

\author{Yasuyuki Sumi}
\authornotemark[1]
\orcid{0000-0002-9247-4208}
\affiliation{%
  \institution{Future University, Hakodate}
  \city{Hakodate}
  \state{Hokkaido}
  \country{Japan}
}
\email{sumi@acm.org}

\renewcommand{\shortauthors}{Dizon et al.}

\begin{abstract}
Large Language Models (LLMs) are increasingly embedded into applications, allowing users to complete tasks either through direct manipulation or by delegating actions to conversational agents. However, little is known about how users balance these modalities when both are available. We present a web-based content management system augmented with an LLM agent through the Model Context Protocol (MCP), enabling users to perform CRUD tasks through a graphical interface, a conversational agent, or both. We conducted a between-subjects study (N=73) comparing three interaction modes: Traditional-Only, AI-First, and Hybrid. Across sixteen scenarios, we analyzed task completion time, interaction logs, and delegation behavior. AI-assisted interaction significantly reduced clicks, page navigations, and scrolling, indicating lower interaction effort. Surprisingly, these reductions did not translate into faster task completion, as task duration did not differ significantly across conditions. We also found no significant relationship between CRUD operation type and delegation, suggesting that users did not systematically avoid delegating higher-risk actions. Instead, delegation varied far more between participants than between tasks, with individual differences accounting for roughly half the variance in assistant use (ICC = .50). Our findings suggest that the primary benefit of human–agent interfaces may be reducing interaction effort rather than improving speed, and that delegation reflects who the user is more than what the task demands.

\end{abstract}

\ccsdesc[500]{Human-centered computing~Human computer interaction (HCI)}
\ccsdesc[300]{Human-centered computing~Interactive systems and tools}

\keywords{Conversational Agents, Human-Agent Interaction, Model Context Protocol, Interaction Modality}
\begin{teaserfigure}
  \includegraphics[width=\textwidth]{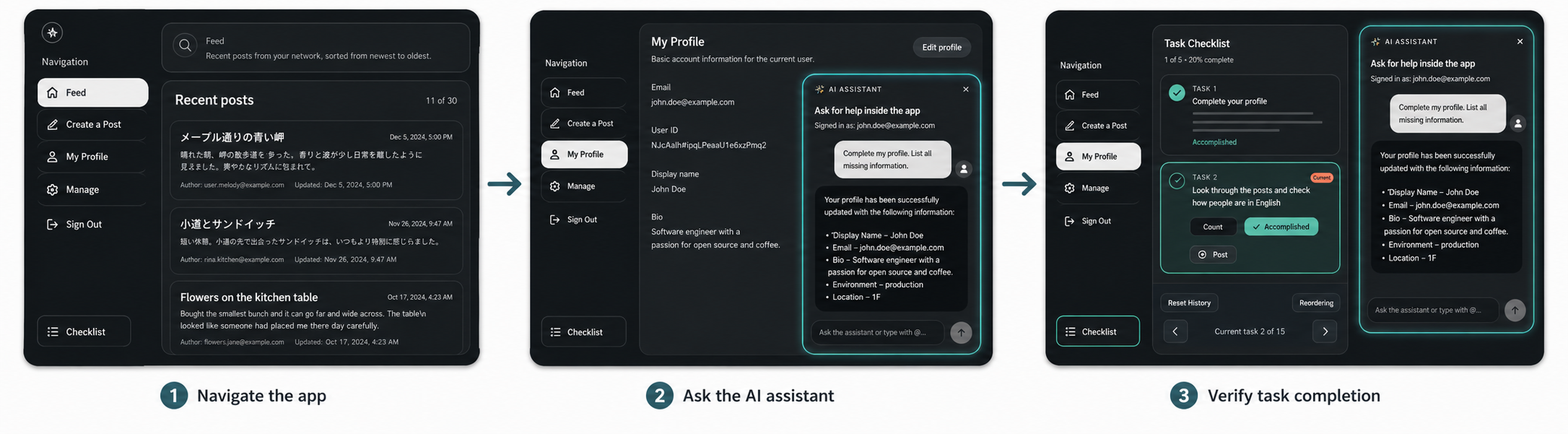}
\caption{Overview of delegAgents. Users perform tasks directly through the graphical interface (1), delegate them to the embedded AI assistant via the MCP-connected backend (2), and verify progress on the task checklist (3).}
  \label{fig:teaser}
\end{teaserfigure}

\received{05 June 2026}
\received[accepted]{01 August 2026}
\received[revised]{16 August 2026}
\received[revised]{24 August 2026}

\maketitle

\newcommand{\hai}{\(HAI\)}
\newcommand{\hci}{\(HCI\)}
\newcommand{\ete}{ETE}

\section{Introduction}

\noindent Over the past years, Large Language Models (LLMs) have significantly evolved. Early iterations, such as GPT-2 and BERT, demonstrated promising capabilities in natural language understanding but remained largely confined to research settings, limited by computational constraints and generalization \cite{llm_for_info_retrieval, devlin-etal-2019-bert}. The introduction of transformer-based architectures and large-scale pretraining on web-scale corpora started a change in capability, with models like GPT-4, Sonnet, and Gemini that exhibit fluent reasoning, instruction-following, and multi-turn dialogue across diverse domains \cite{transformer}. This trajectory has accelerated the transition of LLMs from specialized tools to general-purpose assistants embedded in everyday digital life.

\noindent The rapid growth of LLM-powered chat applications has changed how people interact with systems and applications. Platforms such as ChatGPT, Claude, and Gemini have drawn hundreds of millions of users, with adoption spanning a wide range of everyday tasks including writing assistance, information retrieval, code generation, and decision support \cite{ChatGPT_increase}. Beyond passive querying, users increasingly engage these systems as \textit{active collaborators}, delegating cognitive work that was previously performed through direct exploration and manipulation of traditional user interfaces (UI). This shift in usage patterns raises a question for human-computer interaction (HCI), specifically, human-agent interaction: as users move from exploration and manipulation to conversation and delegation, how does this shape their behavior within task-oriented applications?

\noindent One observable consequence of this shift is in how people approach \textit{exploratory task execution} (\ete), specifically, the process by which users simultaneously discover, navigate, and act within an application to accomplish a defined goal. Unlike purely information-seeking behavior, \ete \hspace{0.5mm} involves a cycle of orientation, targeted action, and verification, where the user must both locate the relevant data and perform an operation on it \cite{Pirolli_2009, exploratory_search}.
Prior work has examined how interface design shapes exploratory behavior in web and database contexts \cite{exploratory_search}, but the rise of conversational LLM interfaces introduces a new dynamic: users may now articulate tasks in natural language, offloading both the navigation and execution steps to an agent. The question still remains whether this changes the nature of exploratory behavior and whether different interaction modalities lead to different task execution patterns.

Aside from the rapid growth of LLMs, a second enabler of this shift is the Model Context Protocol (MCP), an open standard introduced by Anthropic that lets LLM clients discover and invoke functions exposed by external systems \cite{mcp_intro}. Its consequence is that agents move beyond generating content and into acting on a user's behalf. Because most modern web applications already expose their functionality through REST APIs \cite{rest_api_usage}, wrapping those endpoints with an MCP server makes existing functionality available to an agent without modifying the underlying system. The result is a dual-interface scenario: one application reachable through both a graphical interface and a conversational agent, each acting on the same data through the same API layer. A user who once navigated a form to create a record can now simply say so — and must decide, each time, which route to take.

\noindent In this study, we investigate how interaction modality shapes user behavior when performing everyday data tasks in an LLM-augmented web application. Building on the dual-interface scenario described above, we conducted a between-subjects study comparing three interaction modalities: a traditional condition using only the graphical interface, a conversational condition using only the MCP-connected chat agent, and a hybrid condition offering both. By capturing rich behavioral traces across these conditions, we examine how users navigate, delegate, and complete tasks under each paradigm. We focus in particular on the hybrid condition, where users freely chose between direct manipulation and conversational delegation, revealing how people negotiate the boundary between their own agency and that of the agent.

\noindent This study makes the following contributions:
\begin{itemize}
    \item  Present a generalizable architecture for augmenting existing REST-based web applications with conversational agent capabilities through the MCP, requiring minimal modification to the underlying system
    \item  Report findings from a between-subjects study (\textit{N} = \texttt{73}) comparing traditional, AI-First, and hybrid interaction modalities for everyday CRUD tasks, contributing empirical evidence on how interaction modality shapes \ete \hspace{0.5mm} behavior
    \item Characterize delegation behavior in the assistant-enabled conditions, showing that delegation varies far more across users than across tasks, and discuss the implications for designing LLM-augmented applications that appropriately balance user agency and agent autonomy   
\end{itemize}

\section{Related Work}

Our study proposes to analyze the underlying interaction modalities brought about by LLM-Augmented Applications to do \ete. As such, we focus on four (4) domains in relation to our work: (1) Conversational and Graphical Interfaces for Task Execution, (2) Exploratory Task Execution and Interaction Behavior, (3) Task Delegation to LLM Agents, and (4) The Model Context Protocol.

\subsection{Conversational versus Graphical Interfaces for Task Execution}




\noindent The comparison between conversational user interfaces (CUIs) and graphical user interfaces (GUIs) has had a rich history, though most prior work predates
the current generation of LLM-powered agents \cite{McTear2002,Androutsopoulos1995, VNLD_Survey}. A recurring finding, however, is that conversational interfaces lower the learning effort required of users and afford a more natural interaction style, yet do not uniformly outperform graphical interfaces across task types \cite{liu2024conversational}. In repeated-use decision-making tasks, for instance, the initial appeal of natural-language interaction can diminish as users become proficient with the structure of a graphical interface \cite{liu2024conversational}. This suggests that interaction modality interacts with task characteristics in ways that resist simple "better or worse" conclusions — a tension our three-condition design is positioned to examine.


\noindent Another study compared a text-based conversational interface against a conventional web-based GUI for the same decision support system, using a between-subjects design that also manipulated system accuracy \cite{cui_decision_support_system}. Their setup was a single backend made accessible through two distinct interfaces. They studied housing recommendations rather than agentic task execution. Moreover, the conversational interface was not powered by an LLM. However, more recent studies have already examined conversational agents now powered by LLMs \cite{conv_agents_stats_anal, voice_cms}. One study evaluated the effectiveness of conversational agents in performing statistical tasks using GUI-based tools versus conversational agents \cite{conv_agents_stats_anal}. Results from this study indicated that conversational agents outperform a traditional GUI statistical software in all assessed quantitative and qualitative metrics. In another study focusing on voice-driven content management, they found that conversational input can match graphical input in quality even for complex tasks, while pointing towards hybrid interfaces as promising direction \cite{voice_cms}. 

\noindent Our work extends on these studies by analyzing not only conversational and traditional modalities in isolation, but also a hybrid condition in which users freely choose between which they are more comfortable to do a specific task.

\subsection{Exploratory Task Execution and Interaction Behavior}


\noindent Our framing of \ete \hspace{0.5mm} draws on the exploratory search literature,
which distinguishes goal-directed lookup from the more open-ended, iterative process of
discovering and making sense of an information space \cite{exploratory_search}. Although it centers on information foraging rather than data manipulation, its methodological emphasis is on capturing fine-grained behavioral traces such as query reformulations, click streams, and navigation sequences. These traces are valuable because they expose how the interface itself participates in exploration rather than merely hosting it. In a study regarding interactive intent modeling, they demonstrate that when a system actively models and responds to a user's shifting intent, it reshapes not only what users ultimately find but the path they take to get there \cite{interactive_intent_modeling}. We adopt the same methodological stance, treating interaction traces as a primary window into user behavior rather than relying on self-report alone. 

\noindent The intersection of exploratory behavior and conversational interaction is only beginning to be studied. Recent work examining conversational exploratory search collected rich behavioral data which includes querying, clicking, and eye-fixation sequences. These were collected to understand how task complexity and domain expertise shape user behavior in conversational settings \cite{exploratory2025tois}.

\noindent We build on this behavioral-trace tradition but shift the focus from \textit{information foraging} to task execution: in our setting, users must not only locate relevant data but also act upon it through create, read, update, and delete (CRUD) operations. This combination of exploration and consequential action distinguishes our study from prior exploratory search work. 

\subsection{Task Delegation to LLM Agents}




\noindent When an LLM agent can act on external systems, using it is itself an act of delegation. Prior work had established a framework of task delegability built on four factors that shape willingness to delegate to AI: motivation, difficulty, risk, and trust \cite{task_deg_framework}. Particularly, people have a strong preference for machine-in-the-loop designs, rather than allowing AI to have full control, with trust emerging as the factor most strongly correlated with delegation preferences. Following the said framework is crucial in our study as CRUD operations vary naturally in risk. While reading operations are low-consequence and reversible, operations like deletion are destructive and difficult to undo.

\noindent A rather more recent empirical work has applied the said framework towards LLM Agents \cite{plan_then_execute}. Particularly, the study focused on users delegating everyday tasks (such as booking flight tickets or paying via credit card) to an LLM agent under a \textit{plan-then-execute} paradigm, examining how trust unfold when users can inspect and intervene in the agent's actions. Their findings suggested that the value of LLM agents is conditional. While they succeed when a high-quality plan is in place and users stay involved in the execution, trust diminishes easily when a plan only seems plausible on the surface. Another complementary work argued for treating delegation not as an opaque default but as a visible, negotiable decision between user and agent \cite{task_aware_deleg_cues}.

\noindent We extend this work by capturing delegation behavior in practice rather than in self-report. Because our hybrid condition lets users either perform an operation directly or hand it to the agent, we can identify when delegation actually occurs.

\subsection{Model Context Protocol}



\noindent One important component in making LLMs agentic is the Model Context Protocol (MCP). MCP, an open standard introduced by Anthropic, defines a unified, bidirectional interface between LLM and external tools and resources \cite{mcp_intro}. MCP organizes interaction around a host–client–server architecture in
which a server exposes discrete tools that an LLM client can discover and invoke. While the previous study \cite{mcp_intro} focuses primarily on MCPs architecture, adoption landscape, and security implications, the protocol's significance for HCI lies in how readily it can make existing software functionality available to conversational agents.  Recent engineering work demonstrates that conventional REST APIs can be converted into MCP servers with minimal effort, effectively making existing web services "agent-ready" without altering their underlying implementation \cite{rest_to_mcp}.
Recent work has increasingly framed agent interaction not only as conversation, but as orchestration across multiple capabilities and actors \cite{schombs2025conversation}.
This perspective is relevant to our MCP-based implementation, where the assistant does not merely generate responses but invokes application functions on behalf of users.

\section{Model Context Protocol and delegAgents}
\subsection{delegAgents Web Application}
\noindent In order to answer the questions regarding how people interact with LLM-Augmented applications and how CRUD operations are delegated (or not) to these chat assistants, we first developed a system capable of doing these.  

\noindent A \textit{Content Management System} (CMS) web application was developed that features common functionalities that a CMS has: \textit{profile}, \textit{manage user}, \textit{manage post}, and \textit{user role assignment}. We called this CMS, \textit{delegAgents}. \autoref{fig:teaser} shows a portion of the CMS. For the application to read, access, and store data, we also developed a REST API. This REST API is patterned with the features present in the CMS web application. For every entity or model, a complete CRUD counterpart was developed. However, unlike the traditional CMS, the REST API is connected in two ways. First, it is connected through traditional means via pages \& form interactions (e.g. reading data from users, retrieving profile information, submitting a post). Aside from this, it is connected via the MCP through a chat assistant that users are able to interact with. Simply put, the MCP contains tools that pattern an endpoint which the user can call through natural language. Notably, this was also the application used for the experiment.

\subsection{MCP as an Integration Bridge}
\noindent The chat assistant communicated with the CMS web application's backend through the MCP. Rather than re-implementing application logic for the assistant, we used MCP as a bridge between the CMS web application and the REST API. This design was deliberate: by exposing the existing REST endpoints through MCP with minimal modification, we sought to demonstrate the protocol's effectiveness in adapting already-deployed REST APIs to conversational, tool-using agents without substantial re-engineering.

\noindent For the participants' natural language to be processed by the MCP, the AI chat assistant was powered by Gemini 3.1 Flash-Lite. Latency, cost-effectiveness, and intelligence are three main considerations as to why this model was chosen. This model interpreted participants' natural language queries, selected the appropriate MCP tool calls, and invoked the REST API associated via the MCP bridge. The model's responses to the participant were generated on the basis of the results returned from these tool calls. 

\subsection{Redirect Augmentation for Visual Feedback} 
\label{sec:redirect_augmentation}

\noindent Although the MCP in itself is already a powerful tool, we introduced one addition to its standard tool-calling flow. For every applicable tool, we associated a redirect link specifying the location within the web application to which the interface should navigate following a successful execution of that tool call. For example, after a tool that updated a user's profile completed successfully, the application automatically navigated to that profile page.

\noindent The motivation for this augmentation was to provide participants with explicit visual confirmation of the changes effected by the assistant, rather than requiring them to rely solely on the assistant's textual response. By surfacing the resulting application state, the redirect allowed participants to directly verify that the requested change had in fact been applied and that it matched their intent. This decision is grounded in well-established principles of interface feedback \cite{design_of_everyday_things, designing_the_user_interface}. Particularly, the importance of keeping users informed about system status was one of our main motivations for this.




\section{User Study}

\subsection{Participants and Study Design}

\noindent We recruited \textit{N} = 73 undergraduate students, all of whom reported English as their native language. The study employed a between-subject design with three conditions: 1. \textit{Traditional-Only}, 2. \textit{Hybrid}, and 3. \textit{AI-First}. In the \textit{Traditional-Only} condition, participants completed the tasks by navigating the web application directly, with no conversational assistance.  In the Hybrid condition, participants used the web application in the usual manner but were additionally provided with an AI Chat Assistant, and received no directive about which mode to prioritize. Finally, in the \textit{AI-First} condition, participants were explicitly instructed to rely on the AI Chat Assistant as much as possible, while still being permitted to operate the web application directly when needed.

\noindent All participants provided informed consent before participating. Additionally, data collection was acquired following the participants' university's ethics review standard.

\subsection{Apparatus and Tasks}
\noindent All tasks were administered through the CMS  application that we developed for this study. We constructed a total of 16 tasks, each associated with one CRUD operation. Task specification include: locating a specific user then deleting it, viewing your profile and updating information, reading through a post and identifying which texts are in English. These 16 tasks were performed by each participant in the same order.

\begin{table}[t]
\caption{Experimental Tasks and Associated CRUD Operations}
\label{tab:tasks-crud}
\small
\setlength{\tabcolsep}{3pt}
\renewcommand{\arraystretch}{1.15}
\begin{tabularx}{\columnwidth}{@{}
>{\raggedright\arraybackslash}p{0.13\columnwidth}
>{\raggedright\arraybackslash}X
>{\raggedright\arraybackslash}p{0.20\columnwidth}
@{}}
\toprule
Task Number & Task Description & Associated CRUD \\
\midrule
1 & Update profile information. & Update \\
2 & Review posts by language. & Read \\
3 & Create a new post. & Create \\
4 & Find and delete a post. & Delete \\
5 & Edit an existing post. & Update \\
6 & Create another post. & Create \\
7 & Delete an existing post. & Delete \\
8 & View available roles. & Read \\
9 & Inspect role permissions. & Read \\
10 & Create a new role. & Create \\
11 & View system users. & Read \\
12 & Find and update a user. & Update \\
13 & Find and delete a user. & Delete \\
14 & Search and count users by nationality. & Read \\
15 & Create a new user with an assigned role. & Create \\
16 & Verify newly created user details. & Read \\
\bottomrule
\end{tabularx}
\end{table}

\noindent Condition assignment was performed at registration using a sequential \textit{round-robin} scheme: the first registrant was assigned to \textit{Hybrid} (A) condition, the second to \textit{AI-First} (B), the third to the \textit{Traditional Only} (C) and fourth going back to the \textit{Hybrid} condition, continuing in this rotation for all subsequent registrations. This approximated balanced allocation without requiring participants to enroll simultaneously. However, because assignment occurred at registration rather than at completion, the final group sizes were unequal. Originally, there were 78 participants, 26 for each condition. However, because five (5) of the participants did not complete or even start the tasks, attrition was distributed unevenly across conditions. Having said this, the total number of participants for subgroups are as follows: \textit{Hybrid} = 25, \textit{AI-First} = 25, and \textit{Traditional Only} = 23. 
In both assistant-enabled conditions, participants were told that the AI Chat Assistant was capable of completing the assigned tasks. They were not, however, instructed to treat its outputs as guaranteed correct, nor were they directed to verify each action; the redirect-based visual confirmation described in Section~\ref{sec:redirect_augmentation} was available should they choose to check the resulting application state.

\subsection{Measures}
\noindent All participant activity was logged automatically by the web application. As an index of how heavily participants engaged with the interface through direct manipulation, we recorded the number of \textit{UI interaction events}, comprising scrolls, clicks, and page navigations. We also recorded how long it took for our participant to finish each task. This was made possible by tracking both the associated REST API endpoints needed for certain tasks. Task timing began when the participant clicked the \textit{Start Record} button in the CMS, which started Task 01. A task ended when the correct REST endpoint was called either by the CMS itself or the chat assistant (or when they opt to give up by clicking the \textit{fail button} near the task item), at which point its end time was logged and the subsequent task started automatically. Because tasks were chained in this way, each task's measured duration included the full interaction window: for AI-delegated tasks, this encompassed request formulation, AI response latency, and verification of the resulting application state. For the two conditions with assistant access (Hybrid and AI-First), we additionally logged the number of conversational turns or chat interactions exchanged between the participant and the AI Chat Assistant. We treat chat interaction count as a behavioral proxy for assistant reliance, not as a direct measure of trust or delegation quality.

\subsection{Results}
\subsubsection{Task Execution Time}
\label{ssec:res_task_exec_time}

\par We analyzed participants' \textit{task execution time} across the three interaction modes using a linear mixed model on log-transformed duration. Participants in the AI-First condition had the lowest mean task execution time ($\mu=46.60$ seconds), followed by Traditional ($\mu=48.20$ seconds), with the Hybrid condition having the highest mean task execution time ($\mu=51.90$ seconds). A linear mixed model revealed no significant effect of interaction mode on task execution time, $F(2,64.6)=0.26$, $p=.772$.

\subsubsection{UI Event Interactions}
\label{sec:ui_event_interaction}

\begin{figure*}[t]
    \centering
    
    \begin{subfigure}[t]{0.32\linewidth}
        \centering
        \includegraphics[width=\linewidth]{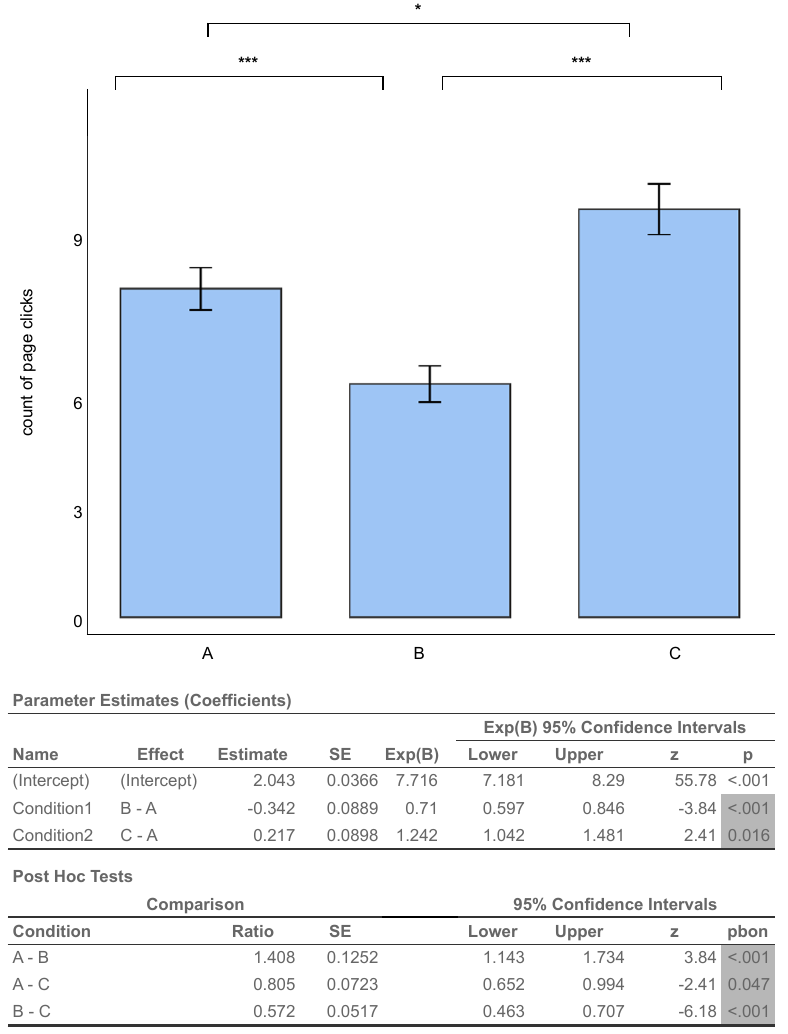}
        \caption{Number of Page Clicks}
        \label{fig:a}
    \end{subfigure}
    \hfill
    \begin{subfigure}[t]{0.32\linewidth}
        \centering
        \includegraphics[width=\linewidth]{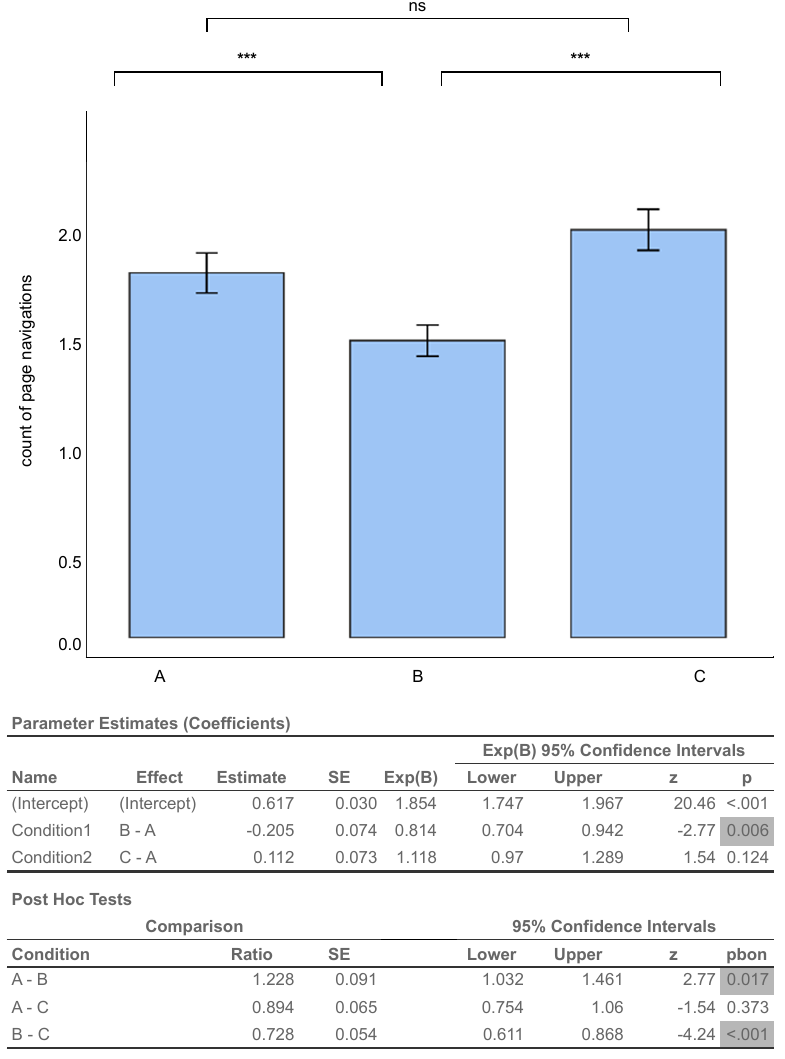}
        \caption{Number of Page Navigations}
        \label{fig:b}
    \end{subfigure}
    \hfill
    \begin{subfigure}[t]{0.32\linewidth}
        \centering
        \includegraphics[width=\linewidth]{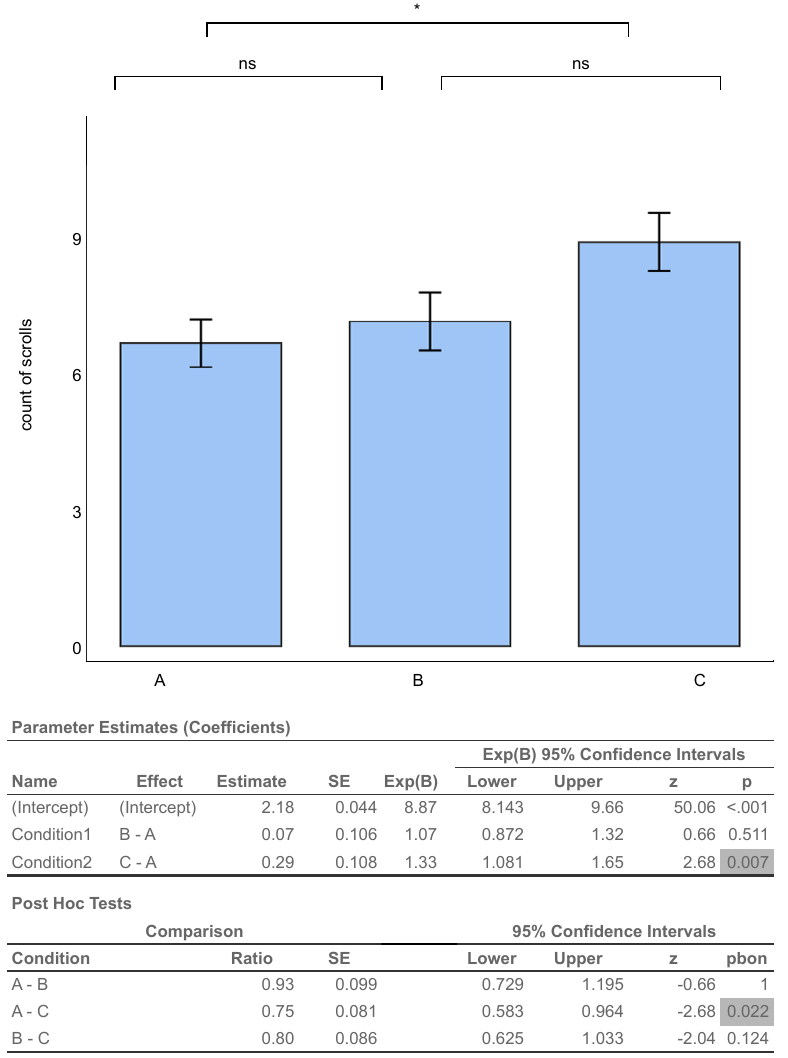}
        \caption{Number of Page Scrolls}
        \label{fig:c}
    \end{subfigure}
    
    \captionsetup{justification=centering}
    \caption{Comparison of \textit{UI Event Interactions} between  different Conditions \\ Conditions are mapped as follows: A - \textit{Hybrid}, B - \textit{AI-First}, \& C \textit{Traditional}}
    \label{fig:conditions}
\end{figure*}

\par We analyzed participants' \textit{UI event interactions} across three measures: page navigations, clicks, and scrolls using Negative Binomial generalized linear model. \autoref{fig:conditions} presents the estimated marginal means across each interaction mode.



\par \textit{Page Navigations:} Participants in the AI-First condition had the lowest mean number of page navigations ($\mu=1.56$), followed by the Hybrid condition ($\mu=1.91$), with the Traditional condition having the highest mean number of page navigations ($\mu=2.14$). A Negative Binomial generalized linear model revealed a significant effect of interaction mode on page navigations \textit{NB-GLM}, $\chi^2(2)=18.30$, $p<.001$, and the omnibus test also revealed a significant effect of interaction mode, $\chi^2(2)=18.50$, $p<.001$. Post-hoc pairwise comparisons indicated significant differences between Hybrid and AI-First ($p=.017$), and between AI-First and Traditional ($p<.001$). No significant difference was observed between Hybrid and Traditional ($p=.373$).


\par \textit{Clicks:} Participants in the AI-First condition had the lowest mean number of clicks ($\mu=5.71$), followed by the Hybrid condition ($\mu=8.05$), with the Traditional condition having the highest mean number of clicks ($\mu=9.99$). A Negative Binomial generalized linear model revealed a significant effect of interaction mode on clicks, \textit{NB-GLM}, $\chi^2(2)=37.50$, $p<.001$, and the omnibus test also revealed a significant effect of interaction mode, $\chi^2(2)=38.20$, $p<.001$. Post-hoc pairwise comparisons indicated significant differences between Hybrid and AI-First ($p<.001$), Hybrid and Traditional ($p=.047$), and AI-First and Traditional ($p<.001$).




\par \textit{Scrolls:} Participants in the Hybrid condition had the lowest mean number of scrolls ($\mu=7.87$), followed by the AI-First condition ($\mu=8.44$), with the Traditional condition having the highest mean number of scrolls ($\mu=10.50$). A Negative Binomial generalized linear model revealed a significant effect of interaction mode on scrolls, (\textit{NB-GLM}, $\chi^2(2)=7.91$, $p=.019$), and the omnibus test also revealed a significant effect of interaction mode, $\chi^2(2)=7.93$, $p=.019$. Post-hoc pairwise comparisons indicated a significant difference between Hybrid and Traditional ($p=.022$). No significant difference was observed between Hybrid and AI-First ($p=1.000$) or between AI-First and Traditional ($p=.124$).

\subsubsection{CRUD Operations and Chat Interactions}
\label{sec:crud_x_chat}

\begin{figure*}[]
   \includegraphics[width=0.75\textwidth]{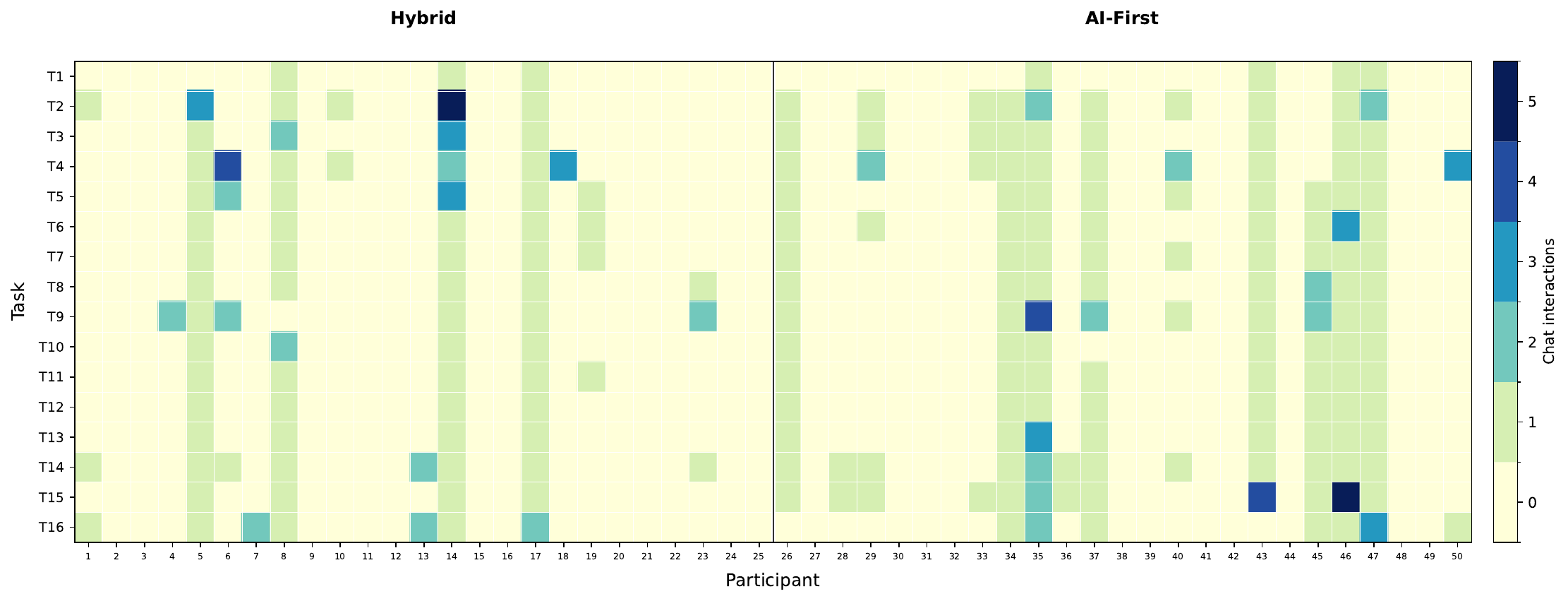} 
   \caption{Chat delegation is a property of the person, not the task. Each box gives the number of chat interactions for one participant (columns 1–50) on one task (rows T1–T16), for the two assistant-enabled conditions (Hybrid, left; AI-First, right)}
    \label{fig:task_interaction_heatmap}
\end{figure*}

\par Because tasks were completed one at a time in a fixed order, each chat turn was attributed to the task the participant was performing at that moment, and thus to that task's CRUD operation. We analyzed the number of chat interactions across CRUD operations using a Poisson generalized mixed model. Read operations had the highest mean number of chat interactions ($\mu=0.391$), followed by Delete operations ($\mu=0.386$), Create operations ($\mu=0.354$), and Update operations ($\mu=0.250$). The model result was not significant, $\chi^2(3)=6.60$, $p=.086$.

\par Beyond the non-significant effect of operation type, the model revealed substantial between-participant variation. The by-participant random intercept accounted for roughly half of the variance in chat interactions (latent-scale $ICC=.50$), and this estimate was essentially unchanged after adding interaction condition as a covariate ($ICC = .50$). Figure~\ref{fig:task_interaction_heatmap} shows that variation runs across participants (columns) rather than tasks (rows). 
\section{Discussion}

\par We frame our discussion based on three guiding questions. We look at how users interact with the agents and how their delegation with it affects the overall interactions. 


\subsection{Reconsidering Risk-Calibrated Delegation}
\label{sec:risk-calibrated-delegation}

\begin{formal}
\textbf{Question 1: How does operation type influence delegation to the chat assistant ? } 
\end{formal}
\noindent We treated the four CRUD operations as a scale of risk: reading and creating records are fairly low-stakes actions, while updating and deleting carry more weight and can undo or destroy data. Drawing on trust-in-automation theory \cite{lee2004trust}, we expected participants to match their chat interactions to this scale—relying on it freely for low-risk operations and using it less as the risk of doing damage increased. The idea that people pull back from the assistant on riskier operations predicts a clear link between an operation's destructiveness and how much users avoid it \cite{lee2004trust}. However, our data show little sign of that link as seen in Section~\ref{sec:crud_x_chat}. Chat interactions did not track the risk ordering of the operations; in fact, \textit{DELETE}, the most destructive operation, drew descriptively more chat interactions ($\mu=0.386$) than \textit{UPDATE} ($\mu=0.250$), the opposite of what a risk-calibration account predicts.

One possibility is that simply counting how often people used the chat is a coarse measure to capture what trust-in-automation theory is really about. People may handle risk by using the assistant more carefully, not less often. As such, a simple count of interactions cannot detect that. The \textit{task delegability framework} \cite{task_deg_framework} makes a similar point: risk is only one of four factors that influence whether a user delegates a task to a machine, and of those four it is \textit{trust}, not \textit{risk}, that most strongly predicts how people choose to delegate.

\noindent A second finding reinforces prior work: participant-level variability. Chat interactions were shaped more by who the user was than by what the task was as seen in \autoref{fig:task_interaction_heatmap}. Our data locate this variability in the person but cannot say what property of the person produces it; the account below is an interpretation, not a result. Prior studies have established that automation use is driven substantially by stable, person-level dispositions \cite{merritt_ilgen, tia_empirical_evidence_infleunce_trust, huangHai}. Merritt and Ilgen \cite{merritt_ilgen} showed that a person's general tendency to trust machines predicts how much they use automation, even after accounting for the system's actual qualities. Since then, this personal tendency, often called \textit{dispositional trust}, has been treated as a key factor shaping how people interact with automation \cite{tia_empirical_evidence_infleunce_trust, lee2004trust}. Our result is compatible with this: reliance on a conversational agent looks less like a response evoked by a particular task and more like the expression of a relatively stable individual tendency that users bring with them into the system. On this reading, the chat assistant is not a tool whose use is driven by task demands, but one whose adoption reflects a disposition that varies from person to person. 
We therefore treat dispositional trust as a hypothesis rather than a conclusion. Testing it would require pairing this behavioral logging with a propensity-to-trust instrument administered before exposure to the system.

\subsection{Task Duration Across Interaction Modes}
\label{sec:task_dur}

\begin{formal}
\textbf{Question 2: How does the use of a chat assistant affect task execution time?} 
\end{formal}

\noindent One focus was to determine whether a conversational interface is faster or slower than a traditional graphical one. Given our results in Section~\ref{ssec:res_task_exec_time}, this remains unsettled. There is evidence that conversation can reduce navigation effort and friction, especially for users who do not know where to begin. In a randomized controlled comparison, a conversational search interface helped people with low literacy find health information that they could not locate using a conventional keyword-and-form search engine \cite{Bickmore2016}. The non-significance, however, aligns to the argument that no interaction mode is considered generally the fastest. Descriptively, \textit{AI-First} had the lowest mean completion time or fastest task execution time ($\mu=46.6$ seconds). This opposes the narrative that conversation is slower, which is claimed by form-comparison studies \cite{Soni2022}.

\noindent Since the study focuses on comparing different interaction modes: conversational, hybrid, or traditional interface, the finding is reassuring. \textit{AI-First} and \textit{Hybrid access} did not cost users any measurable time compared to the traditional interface, so speed is not a reason to avoid either one. Earlier work found conversation slower but still preferred \cite{Soni2022}. Our findings do not reproduce the speed difference. 


\subsection{UI Interaction Events Across Interaction Modes}
\label{sec:ui_interaction}

\begin{formal}
\textbf{Question 3: How does the chat assistant affect UI interaction events? } 
\end{formal}

The third set of analyses asked whether the interaction mode a participant worked in \textit{Hybrid} (A), \textit{AI-First} (B), or \textit{Traditional} (C), shaped how much direct manipulation of the web application they performed, indexed by the three UI interaction events: \textit{clicks}, \textit{page navigations}, and \textit{scrolls}.  Interaction mode significantly affected all three as seen in Section~\ref{sec:ui_event_interaction}. The effect ran consistently in one direction: AI-First produced the fewest direct actions, Traditional the most, with Hybrid between them on every measure.


\noindent These behavioral-trace results fit within prior comparisons of conversational and graphical interaction. The drop in clicks and page navigations under AI-First is in line with the study which concluded  that LLM agents let users reach their goals through natural conversation and so reduce direct, manual control over function-specific tools; our logs capture that shift directly, as fewer separate manipulation and navigation events \cite{plan_then_execute}. At the same time, the assistant did not remove direct manipulation, and scrolling did not change across modes. This pattern matches work showing that conversational interfaces lower interaction effort but cannot fully replace graphical ones and do not always outperform them \citep{liu2024conversational}. The fact that scrolling persists is consistent with behavioral-trace studies of conversational exploratory search, where scanning and reading remain a large part of the interaction whatever the conversational channel \citep{exploratory2025tois}, which supports our reading that the assistant takes over operation-issuing actions but not the reading that comes before them. On method, where much of this literature relies on task time, accuracy, or self-report \citep{conv_agents_stats_anal}, our objective interaction counts give a more detailed and complementary view of \emph{how} the work shifts across modalities.

\subsection{Chat Interactions (or Lack Thereof) in \textit{AI-First }Condition}
\label{sec:chat_interactions}
A pattern emerges when we consider that the \textit{AI-First} condition explicitly instructed participants to rely on the assistant as much as possible: even under this directive, chat interactions remained sparse and highly uneven. As \autoref{fig:task_interaction_heatmap} shows, delegation in AI-First ranged from participants who invoked the assistant on many tasks to others who used it rarely or not at all, and the instruction lifted average delegation only modestly above the Hybrid condition, where no such directive was given. In other words, telling users to prefer the agent did not translate into
uniform or heavy conversational use. This is consistent with the dispositional account we develop in Section~\ref{sec:risk-calibrated-delegation}; reliance on an assistant
behaves like a relatively stable individual tendency~\cite{merritt_ilgen,
lee2004trust, tia_empirical_evidence_infleunce_trust} that an experimental instruction can nudge but not
override. It also aligns with the \textit{task delegability framework }\cite{task_deg_framework},
in which an explicit prompt to delegate is only one input among several that jointly determine
whether a user actually hands a task to a machine.
Yet the relative absence of chat did not prevent the assistant from reshaping behavior. The AI-First condition still produced the fewest clicks and page navigations of any condition, which appears to contradict its low conversational volume. We read this not as a contradiction but as evidence that delegation, when it did occur, was highly leveraged:
because each successful tool call collapses a multi-step locate-navigate-act sequence into a single conversational turn, even a handful of delegations can remove many discrete manipulation events. The benefit of the assistant in
\textit{AI-First} thus came less from sustained conversation than from occasional, high-impact handoffs. This pattern echoes prior work showing that the value of LLM agents is conditional and that users tend to stay involved rather than
ceding full control \cite{plan_then_execute}. We return to the design implications of this pattern in Section~\ref{sec:design_recommendations}.

\subsection{Design Recommendations}
\label{sec:design_recommendations}

\noindent Our results speak less to whether hybrid interfaces are worthwhile than to how they should be built. Three recommendations follow from the patterns above.

\noindent \textit{Do not make delegation the default}. The AI-First directive was the strongest push toward the assistant our design allowed, and it lifted average delegation only modestly above Hybrid while leaving the spread across participants largely intact (Section~\ref{sec:chat_interactions}). If an explicit instruction cannot produce uniform adoption, an interface default will not either \cite{task_deg_framework, merritt_ilgen}. Systems should make the assistant continuously available and let each user settle at their own level, rather than routing work through it by default and treating direct manipulation as fallback.

\noindent \textit{Optimize the single handoff, not the conversation}. AI-First produced the fewest clicks and page navigations despite sparse chat use, because each successful tool call collapses a locate–navigate–act sequence into one turn (Section~\ref{sec:chat_interactions}). The design target is therefore the cost of a single delegation: how quickly a user can state an intent and confirm the result, not how well the system sustains dialogue. Features that reward extended conversation optimize for behavior most users did not exhibit.

\noindent \textit{Place safeguards on the agent path, not on the user}. Participants did not withhold delegation as operations became more destructive; DELETE drew descriptively more chat interactions than UPDATE (Section~\ref{sec:risk-calibrated-delegation}). Whatever the reason, designers cannot assume users will self-restrict on consequential actions. Confirmation steps, previews of pending changes, and reversible operations belong in the delegated path, and should be at least as strong there as in the graphical one.

\noindent Underlying these is a shift in evaluation. Task duration did not differ across conditions while interaction effort did (Section~\ref{sec:task_dur} and Section~\ref{sec:ui_interaction}), so a hybrid interface judged on speed will appear to do nothing. Effort measures such as clicks, navigations, and steps eliminated per delegation capture what these systems actually change.

\section{Limitations and Future Work}

\noindent Several limitations qualify our findings, and each points toward work that would address it.

\noindent To start, our sample was homogeneous. All participants were undergraduate students who reported English as their native language. This group is younger and more digitally fluent than the general user base of LLM-augmented applications. This matters because prior work finds conversational interfaces most useful for people with low domain or computer literacy \cite{Bickmore2016}, and those users are absent from our sample. Future studies could look into recruiting a more diverse set of participants.

\noindent Second, our delegation measure was coarse. We counted conversational turns as an index of reliance on the assistant. A frequency count cannot separate careful, deliberate use from casual use, so it may hide how users actually manage risk \cite{task_deg_framework}. This is one reason the CRUD result in Section~\ref{sec:risk-calibrated-delegation} should be read as inconclusive rather than as evidence against risk-calibrated delegation. Conversation analysis, think-aloud protocols, and linguistic analysis would suit this question better. Markers such as verification requests, conditional instructions, and hesitation could show whether users deliberate more on high-risk operations even when they delegate at the same rate.

\noindent Third, we collected no self-report data. We logged behavior but did not measure satisfaction, preference, or trust. As noted in Section~\ref{sec:risk-calibrated-delegation}, this is why our dispositional account stays an interpretation rather than a demonstrated mechanism. It also leaves us unable to link the effort reductions we observed to subjective experience. A direct test would pair this behavioral logging with a propensity-to-trust instrument given before participants use the system. 

\noindent Aside from these, condition assignment was imperfect. Assignment happened at registration rather than at completion, and uneven attrition left unequal group sizes. Round-robin allocation approximated balance, but we cannot rule out selection effects. Furthermore, we studied one application and one task order. The domain was a CMS with CRUD operations, and all participants completed the 16 tasks in the same sequence, so practice and fatigue effects are confounded with task position. Our \textit{redirect augmentation} is a further caveat. Each successful tool call navigated the interface automatically, and our logs counted that as one page navigation (Section~\ref{sec:redirect_augmentation}). Some navigation counts in the assistant conditions came from the system, not the participant. This works against our result rather than for it. AI-First had the fewest navigation counts even with these extra events, so the real drop in participant-initiated navigation is larger than our numbers show. The assistant also ran on a single model (Gemini 3.1 Flash-Lite). Model capability, latency, and error behavior likely shape delegation, so our results should not be read as model-independent. Testing across other domains and models would show how far these patterns hold. 

\noindent Lastly, the interface patterns in Section~\ref{sec:design_recommendations} need empirical evaluation. Context-sensitive delegation suggestions, "take over" controls, and transparent execution logs are plausible ways to support movement between doing and delegating, but we did not test them.

\section{Conclusion}
\noindent We investigated how interaction modality shapes user behavior when performing everyday data tasks in an LLM-augmented web application, comparing \textit{Traditional-Only}, \textit{AI-First}, and \textit{Hybrid} conditions in a between-subjects study (\textit{N} = 73) built on a CMS connected to an LLM agent through the Model Context Protocol. Our results show that AI-assisted interaction significantly reduced direct-manipulation effort—fewer clicks and page navigations—with the \textit{AI-First} condition producing the fewest direct actions and the Traditional condition the most, while the Hybrid condition sat between the two without being reliably distinguishable from either. These reductions in interaction effort did not, however, translate into faster task completion: task duration did not differ significantly across conditions. We further found no significant relationship between CRUD operation type and delegation, suggesting that our participants did not systematically withhold delegation for higher-risk actions, and that reliance on the assistant was driven more by who the user was than by what the task demanded.

\noindent Taken together, these findings suggest that the primary benefit of human–agent interfaces lies in reducing interaction effort rather than improving speed, and that individual differences account for more variance in delegation than task characteristics do. For designers, this reframes the choice among conversational, hybrid, and traditional modalities: since AI-First and Hybrid access cost users no measurable time relative to the traditional interface, the basis for that choice should be user experience, preference, and individual fit rather than raw efficiency—and systems should be designed to support fluid movement between direct manipulation and delegation as users negotiate the boundary between their own agency and that of the agent.



\bibliographystyle{ACM-Reference-Format}
\bibliography{base}


\end{document}